\documentclass[12pt, letterpaper]{article}

\usepackage{arxiv}
\usepackage[utf8]{inputenc} 
\usepackage[T1]{fontenc}    
\usepackage{amsmath,amsfonts,amssymb,bbm}
\usepackage{mathtools}
\usepackage{mathptmx}
\usepackage{newtxtext}
\usepackage{newtxmath}
\usepackage{hyperref}
\usepackage{caption}
\usepackage{float}
\usepackage{xcolor}
\usepackage{multirow}
\usepackage{subcaption,booktabs}
\usepackage{rotating}
\usepackage{adjustbox}
\usepackage{graphicx}
\usepackage{array}
\usepackage{environ}
\usepackage[authoryear]{natbib}

\title{The Asymptotic Distribution for a Single Joinpoint Changepoint Model}

\author{
Xueheng Shi \\
Department of Statistics,\\
University of Nebraska-Lincoln \\
\texttt{shixueheng@gmail.com}
\And	
Robert Lund \\
Department of Statistics,\\
University of California, Santa Cruz \\
  \texttt{rolund@ucsc.edu} 
}

\begin{document}

\maketitle

\begin{abstract}
A single joinpoint changepoint model partitions a time series into two segments, joined at the changepoint time by constraining the estimated piecewise linear regression responses to be continuous. This manuscript derives the exact asymptotic distribution of the changepoint existence test statistic gauging whether or not a second segment is necessary.  The identified asymptotic distribution, a supremum of a Gaussian process over the unit interval, is rather unwieldy. The work presented here provides the result and its derivation; quantiles of the asymptotic distribution are presented for the user. This addresses a subtle gap in the changepoint literature.
\end{abstract}

\keywords{Brownian Bridge, Gaussian Process, Joinpoint Model, Supremum.}

\section{Introduction}
\label{intro}

Detecting changepoints, i.e., the locations in time where the statistical properties of a process shift, has become a fundamental statistical task.  Correctly identifying whether shifts occur and their occurrence times is crucial for accurate modeling, interpretation, and forecasting. Changepoint issues now permeate a wide range of disciplines including speech \citep{davis2006structural}, quality control \citep{hawkins2003changepoint}, and climatology \citep{lund2023good}. In some domains, processes exhibit abrupt level shifts without any trend changes (the so-called mean shift paradigm). In other cases, trends can also change. Trend shifts are known to occur in agricultural \citep{Alston2010Persistence}, economic \citep{Brandt2008China}, and climate \citep{beaulieu2024recent} time series.

Misclassifying the nature of a changepoint shift can lead to incorrect conclusions about causality, trends, and future projections.  For example, interpreting an upward mean shift as a trend change may overstate long-term growth; conversely, ignoring a slope change may understate cumulative impacts over time. These distinctions are also critical for counterfactual analysis, policy evaluation, and attribution studies. To understand changes in global temperature trends in the context of natural and anthropogenic influences, we need to consider both mean and trend shifts.  For global series, a discontinuous mean function is physically unrealistic.  This motivates the study of joinpoint (also called joinpin) models: a two-phase linear regression where the regression response is constrained to be continuous.






A gap in the statistical literature exists on the single joinpoint changepoint problem: an asymptotic distribution has not been quantified to date. The joinpoint regression model is used in medicine \citep{kim2022twenty}, biology \citep{gillis2019utility}, and climatology 
\citep{beaulieu2024recent}.  We fill this gap here, deriving an asymptotic distribution for a single joinpoint changepoint test.


\section{Model and Changepoint Test Statistic}

Two or multi-phase regression models where the regression response is required to meet (i.e., be continuous in a discrete sense) between regimes are called joinpoint models.  A two-regime joinpoint regression structure describing this scenario has the form
\begin{equation} 
	\label{eqn:joinpoint}
	X_t = 
	\left\{ 
	\begin{array}{lc}
		\mu + \alpha t + \epsilon_t,                       & \qquad 1  \le t   \leq k,     \\
		\mu + \alpha t + \beta (t-k)+ \epsilon_t,          & \qquad k+1  \le t   \leq n.     \\ 
	\end{array} \right.
\end{equation}
The above formulation ensures continuity in $E[ X_t]$ at the changepoint time $t = k$. The changepoint existence hypothesis is $H_0: \beta = 0$ versus $H_A: \beta \neq 0$, and $k$ is unknown, as are $\mu, \alpha$, and $\beta$. We assume that $\{ \epsilon_t \}$ is independent and identically distributed with mean zero and variance $\sigma^2$.

If time $k$ is known to be the changepoint, then
\[
J_k:=\frac{\hat{\beta}}
{\mbox{Var}(\hat{\beta})^{1/2}}
\]
should be large in absolute magnitude (statistically non-zero).  Here and elsewhere, hats on quantities denote estimators.  Because the time $k$ is unknown, the quantity
\[
J_{\rm max} = \max_{2 \leq k \leq n-2} |J_k|
\]
is used as the test statistic for the existence of a joinpoint changepoint. The point of this paper is to identify the asymptotic distribution of $J_{\rm max}$ as $n \rightarrow \infty$.

Several tactics exist to derive this asymptotic distribution. \cite{Julious2001} provides a closed form for $\hat{\beta}$ via Lagrange multipliers, expressing estimators of $\mu$, $\alpha$, and $\beta$ by relating them to an unconstrained optimization involving two unconstrained linear segments. While interesting, this does not seem to lead to any computational reductions; moreover, it is challenging to identify $\mbox{Var}(\hat{\beta})$ with this setup. Unfortunately, results in \cite{jandhyala1997iterated} and \cite{robbins2016general} do not apply to this case and would apparently take significant work to modify. As such, we take a classic approach to obtain $\hat{\beta}$ and derive its variances and covariances. While the ideas are straightforward, the computations are quite tedious.

A sum of squared errors for $\mu, \alpha$ and $\beta$ when $k$ is known to be the changepoint time is 
\small{
	\begin{align}
		S(\mu, \alpha, \beta)= \min _{\mu, \alpha, \beta} \left[ \sum_{t=1}^k \left(X_t-\mu-\alpha t\right)^2 + \sum_{t=k+1}^n\left(X_t-\mu-\alpha t-\beta(t-k)\right)^2\right].
	\end{align}
}
Taking partial derivatives with respect to $\mu, \alpha$, and $\beta$, respectively and equating the results to zero, we obtain a three dimensional linear system of equations:
\begin{align*}
	\qquad &\sum_{t=1}^{n} X_{t}=n \hat{\mu}+\hat{\alpha} \sum_{t=1}^{n} t+\hat{\beta} \sum_{t=k+1}^{n}(t-k)\\  
	\qquad &\sum_{t=1}^{n} t X_{t}=\hat{\mu} \sum_{t=1}^{n} t+\hat{\alpha} \sum_{t=1}^{n} t^{2}+\hat{\beta} \sum_{t=k+1}^{n}(t-k) t\\
	\qquad &\sum_{t=k+1}^{n}(t-k) X_{t}=\hat{\mu} \sum_{t=k+1}^{n}(t-k)+\hat{\alpha} \sum_{t=k+1}^{n} t(t-k)+\hat{\beta} \sum_{t=k+1}^{n}(t-k)^{2}
\end{align*}
for the estimators $\hat{\mu}, \hat{\alpha}$, and $\hat{\beta}$,

With some computational effort, the estimate for $\beta$ when a changepoint exists at time $k$ is
\begin{align}
	\hat{\beta}_k=\frac{(n V_{3}-b V_{1})(n c-a^{2})-(n V_{2}-a V_{1})(n d-a b)}{(n e-b^{2})(n c-a^{2})-(n d-a b)^{2}},    
\end{align}
where
\[
a=\sum_{t=1}^n t= \frac{n(n+1)}{2}, \quad
b=\sum_{t=k+1}^n(t-k)=\frac{(n-k)(n-k+1)}{2},
\]
\[ 
c=\sum_{t=1}^n t^2=\frac{n(n+1)(2 n+1)}{6}, 
\quad
d = \sum_{t=k+1}^n(t-k) t=\frac{(n-k)(n-k+1)(2n+k+1)}{6}, 
\]
and
\[
e = \sum_{t=k+1}^n(t-k)^2=\frac{(n-k)(n-k+1)(2(n-k)+1)}{6}. 
\]
Here, 
\[
V_{1}=\sum_{t=1}^{n} X_{t}, \qquad 
V_{2}=\sum_{t=1}^{n} t X_{t}, \qquad 
V_{3}=\sum_{t=k+1}^{n}(t-k) X_{t}   
\]

Expressing $\hat{\beta}_k$ as a linear combination of $V_1, V_2$, and $V_3$ gives 
\begin{align*}
	\hat{\beta}_k & =\frac{(b c-a d) n}{D} V_{1}+\frac{(d n-a b) n}{D} V_{2}+\frac{\left(a^{2}-c n\right) n}{D} V_{3},
\end{align*}
with $D=(ab-dn)^{2}-(b^{2}-ne)(a^{2}-nc)$.  
As the variances and covariances of $V_1, V_2, V_3$ are computable, we find
\begin{equation*}
	\text{Var}(\hat{\beta}_k) = \frac{V_{num}}{V_{dem}},  
\end{equation*}
where
\begin{align*}
	V_{num} &=\frac{1}{864}\left(11 k^{2}-32 k^{3}+33 k^{4}-14 k^{5}+2 k^{6}\right) n^{5}\\
	&+\frac{1}{864}\left(-23 k+84 k^{2}-98 k^{3}+43 k^{4}-6 k^{5}\right) n^{6}\\
	&+\frac{1}{864}\left(12-72 k+91 k^{2}-12 k^{3}-27 k^{4}+14 k^{5}-2 k^{6}\right) n^{7}\\
	&+\frac{1}{864}\left(20-18 k-73 k^{2}+96 k^{3}-43 k^{4}+6 k^{5}\right) n^{8}\\
	&+\frac{1}{864}\left(-8+80 k-102 k^{2}+44 k^{3}-6 k^{4}\right) n^{9}\\
	&+\frac{1}{864}\left(-24+41 k-11 k^{2}+2 k^{3}\right) n^{10}\\
	&+\frac{1}{864}(-4-8 k) n^{11}\\
	&+\frac{1}{216}  n^{12},  
\end{align*}
and
\begin{align*}
	V_{dem}& =\frac{\left(121 k^{4}-704 k^{5}+1750 k^{6}-2420 k^{7}+2029 k^{8}-1052 k^{9}+328 k^{10}-56 k^{11}+4 k^{12}\right) n^{4}}{5184}\\
	&+\frac{\left(-506 k^{3}+3320 k^{4}-9050 k^{5}+13406 k^{6}-11796 k^{7}+6302 k^{8}-1992 k^{9}+340 k^{10}-24 k^{11}\right) n^{5}}{5184}\\
	&+\frac{\left(793 k^{2}-6216 k^{3}+19208 k^{4}-31026 k^{5}+28848 k^{6}-15868 k^{7}+5061 k^{8}-860 k^{9}+60 k^{10}\right) n^{6}}{5184}\\
	&+\frac{\left(-552 k+5768 k^{2}-21322 k^{3}+38490 k^{4}-38010 k^{5}+21350 k^{6}-6788 k^{7}+1144 k^{8}-80 k^{9}\right) n^{7}}{5184}\\
	&+\frac{\left(144-2648 k+12966 k^{2}-27138 k^{3}+28332 k^{4}-15682 k^{5}+4782 k^{6}-800 k^{7}+60 k^{8}\right) n^{8}}{5184}\\
	&+\frac{\left(480-4048 k+10464 k^{2}-10940 k^{3}+4892 k^{4}-1116 k^{5}+196 k^{6}-24 k^{7}\right) n^{9}}{5184}\\
	&+\frac{\left(496-1840 k+1113 k^{2}+1082 k^{3}-715 k^{4}+100 k^{5}+4 k^{6}\right) n^{10}}{5184}\\
	&+\frac{\left(64+568 k-1384 k^{2}+512 k^{3}-80 k^{4}\right) n^{11}}{5184}\\
	&+\frac{\left(-144+392 k-24 k^{2}+16 k^{3}\right) n^{12}}{5184}\\
	&+\frac{(-32-64 k) n^{13}}{5184}\\
	&+\frac{n^{14}}{324}.    
\end{align*}

Some work with the above expressions yields, as $n \rightarrow \infty$, 
\[
\mbox{Var}(\hat{\beta}_k)=\frac{3}{n^3}\frac{\left(\frac{k}{n}\right)^3 \left(1-\frac{k}{n}\right)^3+o(n^{-1})}{\left(\frac{k}{n}\right)^6 \left(1-\frac{k}{n}\right)^6+o(n^{-1})}.
\]
Here, $o(n^{-1})$ is a generic sequence of numbers $r_n$ with $nr_n \rightarrow 0$ as $n \rightarrow \infty$.

To get the asymptotic distribution of $J_{\rm max}$ under the null hypothesis of no changepoints, we note that as $n \rightarrow \infty$, $J_k \stackrel {{\cal D}} \rightarrow N(0,1)$ by the central limit theorem. As such, any limit will be a Gaussian process.  To obtain the covariance function of this limit, we need to identify $\mbox{Cov}(J_k, J_\ell)$ when $1 \leq k \leq \ell \leq n$.  While a tedious endeavor, one can show that 
\begin{equation}
	\label{bigder}
	\mbox{Cov}(J_k, J_\ell) = \frac{C_{num}}{C_{dem}}
\end{equation}
where
\begin{align*}
	C_{num} &=\frac{1}{864}(11 k l-16 k^{2} l+5 k^{3} l-16 k l^{2}+23 k^{2} l^{2}-7 k^{3} l^{2}+5 k l^{3}-7 k^{2} l^{3}+2 k^{3} l^{3}) n^{5} \\
	& +\frac{1}{864}(-11 k+16 k^{2}-5 k^{3}-12 l+52 k l-43 k^{2} l+11 k^{3} l+16 l^{2}-46 k l^{2}+21 k^{2} l^{2}\\
	& \quad-3 k^{3} l^{2}-4 l^{3}+11 k l^{3}-3 k^{2} l^{3}) n^{6} \\
	& +\frac{1}{864}(12-36 k+20 k^{2}-4 k^{3}-36 l+51 k l-2 k^{2} l-5 k^{3} l+20 l^{2}-2 k l^{2}-17 k^{2} l^{2} \\
	& \quad+7 k^{3} l^{2}-4 l^{3}-5 k l^{3}+7 k^{2} l^{3}-2 k^{3} l^{3}) n^{7}\\
	&+\frac{1}{864}(20-10 k-12 k^{2}+6 k^{3}-8 l-49 k l+40 k^{2} l-11 k^{3} l-12 l^{2}+46 k l^{2}-21 k^{2} l^{2} \\
	&\quad+3 k^{3} l^{2}+4 l^{3}-11 k l^{3}+3 k^{2} l^{3}) n^{8} \\
	&+\frac{1}{864}(-8+40 k-20 k^{2}+4 k^{3}+40 l-62 k l+18 k^{2} l-20 l^{2}+18 k l^{2}-6 k^{2} l^{2}+4 l^{3}) n^{9} \\
	&+\frac{1}{864}(-24+21 k-4 k^{2}-k^{3}+20 l-3 k l+3 k^{2} l-4 l^{2}) n^{10} \\
	&+\frac{1}{864}(-4-4 k-4 l) n^{11} \\
	&+\frac{1}{216}n^{12},
\end{align*}

and
\begingroup
\allowdisplaybreaks
\begin{align*}
	C_{dem}&=\frac{1}{5184}(121 k^{2} l^{2}-352 k^{3} l^{2}+363 k^{4} l^{2}-154 k^{5} l^{2}+22 k^{6} l^{2}-352 k^{2} l^{3}+1024 k^{3} l^{3} \\
	\quad&  -1056 k^{4} l^{3}+448 k^{5} l^{3}-64 k^{6} l^{3}+363 k^{2} l^{4}-1056 k^{3} l^{4}+1089 k^{4} l^{4} \\
	\quad&  -462 k^{5} l^{4}+66 k^{6} l^{4}-154 k^{2} l^{5}+448 k^{3} l^{5}-462 k^{4} l^{5}+196 k^{5} l^{5}-28 k^{6} l^{5} \\
	\quad &  +22 k^{2} l^{6}-64 k^{3} l^{6}+66 k^{4} l^{6}-28 k^{5} l^{6}+4 k^{6} l^{6}) n^{4} \\
	& +\frac{1}{5184}(-253 k^{2} l+736 k^{3} l-759 k^{4} l+322 k^{5} l-46 k^{6} l-253 k l^{2}+1848 k^{2} l^{2}-3766 k^{3} l^{2} \\
	\quad& +3245 k^{4} l^{2}-1242 k^{5} l^{2}+168 k^{6} l^{2}+736 k l^{3}-3766 k^{2} l^{3}+6272 k^{3} l^{3} \\
	\quad&  -4610 k^{4} l^{3}+1564 k^{5} l^{3}-196 k^{6} l^{3}-759 k l^{4}+3245 k^{2} l^{4}-4610 k^{3} l^{4} \\
	& \quad +2838 k^{4} l^{4}-800 k^{5} l^{4}+86 k^{6} l^{4}+322 k l^{5}-1242 k^{2} l^{5}+1564 k^{3} l^{5} \\
	\quad&  -800 k^{4} l^{5}+168 k^{5} l^{5}-12 k^{6} l^{5}-46 k l^{6}+168 k^{2} l^{6}-196 k^{3} l^{6}+86 k^{4} l^{6} \\
	\quad&  -12 k^{5} l^{6}) n^{5} \\
	& +\frac{1}{5184}(132 k^{2}-384 k^{3}+396 k^{4}-168 k^{5}+24 k^{6}+529 k l-2724 k^{2} l+4558 k^{3} l \\
	\quad&  -3365 k^{4} l+1146 k^{5} l-144 k^{6} l+132 l^{2}-2724 k l^{2}+9300 k^{2} l^{2} \\
	\quad& -11980 k^{3} l^{2}+7044 k^{4} l^{2}-1932 k^{5} l^{2}+204 k^{6} l^{2}-384 l^{3}+4558 k l^{3} \\
	\quad&  -11980 k^{2} l^{3}+12420 k^{3} l^{3}-5858 k^{4} l^{3}+1204 k^{5} l^{3}-88 k^{6} l^{3}+396 l^{4} \\
	\quad& -3365 k l^{4}+7044 k^{2} l^{4}-5858 k^{3} l^{4}+2245 k^{4} l^{4}-342 k^{5} l^{4}+12 k^{6} l^{4}-168 l^{5} \\
	\quad& +1146 k l^{5}-1932 k^{2} l^{5}+1204 k^{3} l^{5}-342 k^{4} l^{5}+36 k^{5} l^{5}+24 l^{6}-144 k l^{6} \\
	\quad& +204 k^{2} l^{6}-88 k^{3} l^{6}+12 k^{4} l^{6}) n^{6} \\
	& +\frac{1}{5184}(-276 k+1228 k^{2}-1816 k^{3}+1176 k^{4}-352 k^{5}+40 k^{6}-276 l+3312 k l \\
	\quad& -8845 k^{2} l+9380 k^{3} l-4587 k^{4} l+1006 k^{5} l-82 k^{6} l+1228 l^{2}-8845 k l^{2} \\
	\quad& +17378 k^{2} l^{2}-14066 k^{3} l^{2}+5253 k^{4} l^{2}-766 k^{5} l^{2}+22 k^{6} l^{2}-1816 l^{3} \\
	\quad& +9380 k l^{3}-14066 k^{2} l^{3}+8752 k^{3} l^{3}-2546 k^{4} l^{3}+292 k^{5} l^{3}-4 k^{6} l^{3} \\
	\quad& +1176 l^{4}-4587 k l^{4}+5253 k^{2} l^{4}-2546 k^{3} l^{4}+516 k^{4} l^{4}-36 k^{5} l^{4}-352 l^{5} \\
	\quad& +1006 k l^{5}-766 k^{2} l^{5}+292 k^{3} l^{5}-36 k^{4} l^{5}+40 l^{6}-82 k l^{6}+22 k^{2} l^{6} \\
	\quad& -4 k^{3} l^{6} ) n^{7} \\
	& +\frac{1}{5184} (144-1324 k+2948 k^2-2616 k^3+1064 k^4-176 k^5+8 k^6-1324 l+7070 k l \\
	\quad & -10953 k^{2} l+6976 k^{3} l-1931 k^{4} l+134 k^{5} l+16 k^{6} l+2948 l^{2}-10953 k l^{2} \\
	\quad& +12252 k^{2} l^{2}-5734 k^{3} l^{2}+1085 k^{4} l^{2}-66 k^{5} l^{2}-2616 l^{3}+6976 k l^{3} \\
	\quad& -5734 k^{2} l^{3}+2328 k^{3} l^{3}-350 k^{4} l^{3}+12 k^{5} l^{3}+1064 l^{4}-1931 k l^{4} \\
	\quad&  +1085 k^{2} l^{4}-350 k^{3} l^{4}+36 k^{4} l^{4}-176 l^{5}+134 k l^{5}-66 k^{2} l^{5}+12 k^{3} l^{5}+8 l^{6} \\
	\quad&  +16 k l^{6}) n^{8}\\
	& +\frac{1}{5184}(480-2024 k+2464 k^{2}-1168 k^{3}+160 k^{4}+32 k^{5}-8 k^{6}-2024 l+5536 k l \\
	\quad & -4302 k^{2} l+1164 k^{3} l+98 k^{4} l-48 k^{5} l+2464 l^{2}-4302 k l^{2}+2244 k^{2} l^{2} \\
	\quad & -688 k^{3} l^{2}+66 k^{4} l^{2}-1168 l^{3}+1164 k l^{3}-688 k^{2} l^{3}+176 k^{3} l^{3}-12 k^{4} l^{3} \\
	\quad & +160 l^{4}+98 k l^{4}+66 k^{2} l^{4}-12 k^{3} l^{4}+32 l^{5}-48 k l^{5}-8 l^{6}) n^{9} \\
	\quad & +\frac{1}{5184}(496-920 k+292 k^{2}+176 k^{3}-148 k^{4}+24 k^{5}-920 l+529 k l+365 k^{2} l  \\
	\quad & -270 k^{3} l+48 k^{4} l+292 l^{2}+365 k l^{2}+121 k^{2} l^{2}-22 k^{3} l^{2}+176 l^{3}-270 k l^{3} \\
	\quad & -22 k^{2} l^{3}+4 k^{3} l^{3}-148 l^{4}+48 k l^{4}+24 l^{5}) n^{10} \\
	& +\frac{1}{5184}(64+284 k-364 k^{2}+168 k^{3}-24 k^{4}+284 l-656 k l+88 k^{2} l-16 k^{3} l-364 l^{2}\\
	&\quad +88 k l^{2}+168 l^{3}-16 k l^{3}-24 l^{4}) n^{11} \\
	&+\frac{1}{5184}(-144+196 k-44 k^{2}+8 k^{3}+196 l+64 k l-44 l^{2}+8 l^{3}) n^{12} \\
	& +\frac{1}{5184}(-32-32 k-32 l) n^{13} \\
	&+\frac{1}{324}n^{14}.
\end{align*}
\endgroup

\texttt{Mathematica} was used to assist with the calculation of above variance and covariance. Another exercise with the above expressions shows the limit in \ref{bigder} as 
\begin{equation}
	\label{Ozzy}
	\lim_{n \rightarrow \infty} \mbox{Cov}(J_k, J_\ell) = \frac{1}{2} \frac{(3s- t - 2st)}{s(1-t)} 
	\sqrt{\frac{t(1-s)}{s(1-t)}}. 
\end{equation}
Here, $k \leq \ell$ and $n \rightarrow \infty$ in a manner such that $k/n \rightarrow t$ and $\ell/n \rightarrow s$, with $0 < t \leq s < 1$. 

Weak convergence theory on general spaces  \citep{billingsley2013convergence} now essentially shows (the finiteness caveats below) that as $n \rightarrow \infty$, 
\[
J_{\rm max} \stackrel {{\cal D}} {\longrightarrow} \sup_{t \in (0,1)} |G(t)|,
\]
where $\{ G(t) \}_{t=0}^1$ is a zero-mean unit-variance Gaussian process with 
\[
\mbox{Cov}(G(t), G(s)) = \frac{\frac{3}{2}s - \frac{t}{2}-st }{s(1-t)}\sqrt{\frac{t(1-s)}{s(1-t)}}, \qquad 0 < t \leq s < 1. 
\]

Unfortunately, $\sup_{t \in (0,1)} |G(t)| = \infty$ with probability one. This can be seen as follows.   First, it is known that 
\begin{equation}
	\label{div}
	\sup_{ t \in (0,1)} \frac{|B(t)|}{\sqrt{t(1-t)}} = \infty
\end{equation}
with probability one. Here, $\{ B(t) \}_{t=0}^{t=1}$ is a standard Brownian bridge process; viz. 
\[
B(t) = W(t)-tW(1), \quad  t \in [0,1],
\]
where $\{ W(t) \}_{t=0}^{t=1}$ is the standard Brownian motion.  The divergence in (\ref{div}) essentially follows from the divergence 
\[
\sup_{t \in (0, \kappa)} \frac{W(t)}{\sqrt{t}} = \infty
\]
with probability one for every fixed $\kappa > 0$, which in turn is justified from the law of the iterated logarithm for Brownian motion. See \cite{qualls1977law}, \cite{csorgo1997limit}, and \cite{horvath2024change} for further technical detail.

To connect this to the above, one can establish the bound 
\[
\frac{\frac{3}{2}s - \frac{t}{2}-st }{s(1-t)} \leq 3
\]
for any $t,s$ satisfying $0 \leq t \leq s \leq 1$.   Hence, the covariance function on the right hand side of (\ref{Ozzy}) obeys
\[
\frac{1}{2} \frac{(3s- t - 2st)}{s(1-t)} 
\sqrt{\frac{t(1-s)}{s(1-t)}} \leq 
3 \sqrt{\frac{t(1-s)}{s(1-t)}},
\]
which is the covariance function of $\{ \sqrt{3} B_t/\sqrt{t(1-t)} \}$. By a stochastic comparison, it follows that $\sup_{t \in (0,1)} |G(t)| = \infty$
with probability one.  

Given the above, we proceed by truncating the boundaries near zero and unity, using 
\[
J_{\rm max, \delta} := 
\max_{ k: \delta < k/n < 1-\delta } |J_k|
\]
as the changepoint detection statistic. This statistic has the limit
\[
J_{\rm max, \delta} \stackrel {{\cal D}} {\longrightarrow} \sup_{t \in (\delta, 1-\delta)} |G(t)|,
\]
which is finite with probability one.  Asymptotic tables for $J_{\rm max , \delta}$ are reported in Table \ref{tab:scpt-critical-values_trends} for several common truncation values of $\delta$.

\begin{table}[ht]
	\caption{Asymptotic quantiles for the joinpoint changepoint test statistic $J_{{\rm max}, \delta}$.}
	\footnotesize{ 
		\begin{center}
			\begin{tabular}{ c c c c c c }
				\hline
				Statistic & 90.0\% & 95.0\% & 97.5\% & 99.0\% & 99.9\%  \\
				\hline
				$J_{\max, .01}$ & 2.530 &2.795 &3.038 &3.327 &3.964  \\
				$J_{\max, .05}$ & 2.380 &2.658 &2.908 &3.207 &3.852   \\
				$J_{\max, .10}$ &2.285 &2.570 &2.827 &3.132 &3.792    \\
				\hline
			\end{tabular}
			\label{tab:scpt-critical-values_trends}
		\end{center}
	}
\end{table}

\section{Application}

\begin{figure}[H]
	\centering
	\includegraphics[width=0.95\textwidth]{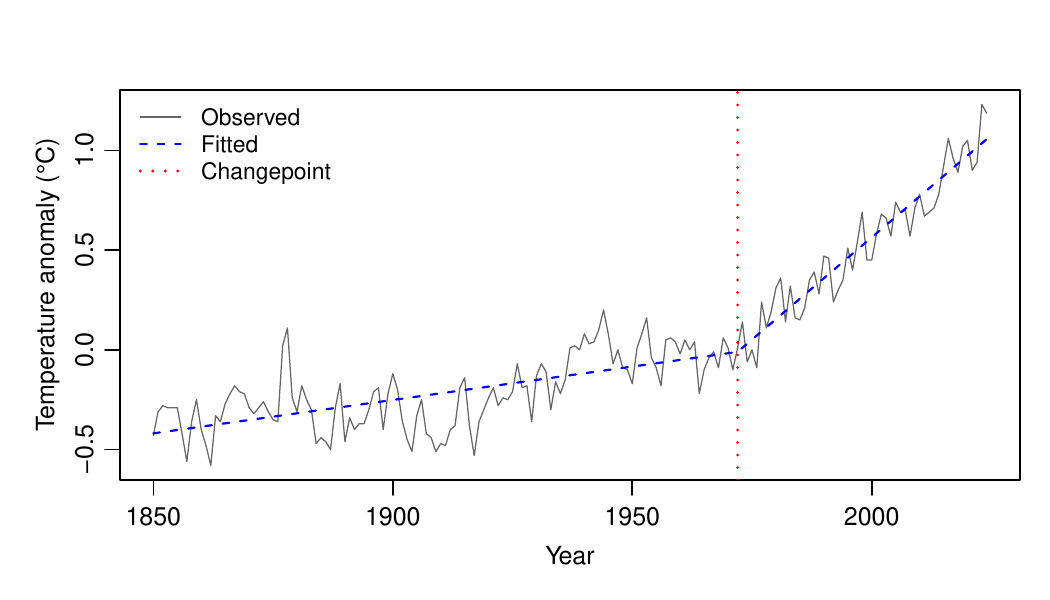}
	\caption{NOAA's annual global temperature anomalies over 1850–2023.}
	\label{fig:Berkeley_Trend_changepoint}
\end{figure}

We now illustrate the methods by examining the merged Land–Ocean Global Surface Temperature record from the National Oceanic and Atmospheric Administration (NOAAGlobalTemp v5.1.0) \citep{Vose_al_2021}. The version analyzed here was compiled by the University of California, Berkeley and is available at \url{https://www.ncei.noaa.gov/access/monitoring/climate-at-a-glance/global/time-series/globe/land_ocean/1/9/1850-2023}. Annual temperature anomalies over the 174-year period 1850–2023 were tested for possible slope changes using our joinpoint model. The data are plotted in Figure~\ref{fig:Berkeley_Trend_changepoint}. As the Earth's temperature is arguably continuous in time, a joinpoint model enforcing continuity of the regression responses between the two regimes seems appropriate.

Table~\ref{tab:amoc_Berkeley} presents the results of the $J_{\rm max}$ slope change test, where the year 1850 is bookkept as $t=1$. The $J_{\rm max}$ statistic peaks in 1972 at 17.46, far above any reasonable critical value.   This same maximum applies to all reasonable truncation levels $\delta$ and suggests a roughly sixfold increase in slope, presumably attributable to anthropogenic global warming. Parameter estimators for the two regimes are reported in the table.  The estimated piecewise trends are superimposed on the data in Figure~\ref{fig:Berkeley_Trend_changepoint}; visually, the fit appears excellent. Table~\ref{tab:amoc_Berkeley} reports the corresponding 95th critical value when $\delta=0.05$; the resulting $p$-values are extremely small (below 0.1\% for any reasonable $\delta$). We also examined the 1970–2023 subperiod for an additional trend shift, but nothing was detected, consistent with the findings of \cite{beaulieu2024recent}.

\begin{table}[H]
	\centering
	\footnotesize
	\setlength{\tabcolsep}{4pt} 
	\caption{Joinpoint test summary with the 95\% quantile when $\delta=0.05$. The $p$-value is less than 0.1\%.}
	\label{tab:amoc_Berkeley}
	\begin{tabular}{@{}lccccccc@{}}
		\toprule
		Test & $\hat{\tau}$ & Test Statistic & 95\% Quantile & Left Intercept & Left Slope & Right Intercept & Right Slope \\
		\midrule
		$J_{\rm max}$ & 1972 & 17.46 & 2.658 & $-6.61$ & 0.0034 & $-40.44$ & 0.0201 \\
		\bottomrule
	\end{tabular}
\end{table}

\section{Conclusion}

This paper derives the asymptotic distribution of the joinpoint test statistic $J_{\rm max}$ for detecting a single changepoint in a continuous two-phase linear regression model. We show that the untrimmed supremum diverges and obtain a well-defined Gaussian-process limit for the trimmed statistic $J_{\rm max,\delta}$, allowing practical critical values to be computed. These results fill a gap in the changepoint literature and provide a simple, rigorous framework for identifying slope changes when continuity of the mean response is required. An application to the 1850--2023 global temperature record illustrates the method, revealing a significant slope increase in the early 1970s with no additional evidence of later trend shifts. The theory developed here may be extended to settings with multiple joinpoints, dependent errors, or additional smoothness constraints.

\section*{Acknowledgments} Robert Lund thanks National Science Foundation Grant DMS-2113592 for partial support; Xueheng Shi thanks University of Nebraska-Lincoln Grant ARD-2162251011 for partial support.

\vspace{1cm}

\section{Data Statement}
Berkeley temperature series analyzed in this paper can be downloaded were listed where they first appeared. \texttt{R} code is available on \url{https://github.com/shixueheng/AMOC}.

%




\newpage

\section{Appendix: the inference of single joinpoint model}

First, we set up the objective function to be optimized:
\begin{align}
	S(\mu, \alpha, \beta)= \min _{\mu, \alpha, \beta}\left[\sum_{t=1}^k\left(X_t-\mu-\alpha t\right)^2+\sum_{t=k+1}^n\left(X_t-\mu-\alpha t-\beta(t-k)\right)^2\right]
\end{align}

By taking partial derivatives with respect to $\mu, \alpha$, and $\beta$ respectively, we get
\begin{align*}
	\frac{\partial S}{\partial \mu}= & \sum_{t=1}^k 2\left(X_t-\mu-\alpha t\right)(-1)+\sum_{t=k+1}^n\left(X_t-\mu-\alpha t-\beta(t-k)\right)(-1) \stackrel{\text { set }}{=} 0 \\
	\Rightarrow \quad & \sum_{t=1}^k\left(X_t-\mu-\alpha t\right)+\sum_{t=k+1}^n\left(X_t-\mu-\alpha t-\beta(t-k)\right)=0 \\
	& \sum_{t=1}^n X_t-n \mu-\alpha \sum_{t=1}^n t-\beta \sum_{t=k+1}^n(t-k)=0 \\
	\Rightarrow \quad & \sum_{t=1}^n X_t=n \mu+\alpha \sum_{t=1}^n t+\beta \sum_{t=k+1}^n(t-k)
\end{align*}

\begin{align*}
	\frac{\partial S}{\partial \alpha}= & \sum_{t=1}^k 2\left(X_t-\mu-\alpha t\right)(-t)+\sum_{t=k+1}^n 2\left(X_t-\mu-\alpha t-\beta(t-k)\right)(-t) \stackrel{\text { set }}{=} 0 \\
	\Rightarrow \quad & \sum_{t=1}^k\left(X_t-\mu-\alpha t\right) t+\sum_{t=k+1}^n\left(X_t-\mu-\alpha t-\beta(t-k)\right) t=0 \\
	& \sum_{t=1}^n X_t-\sum_{t=1}^n \mu t-\sum_{t=1}^n \alpha t^2-\sum_{t=k+1}^n \beta(t-k) t=0\\
	\Rightarrow \quad & \sum_{t=1}^n t X_t=\mu \sum_{t=1}^n t+\alpha \sum_{t=1}^n t^2+\beta \sum_{t=k+1}^n(t-k) t
\end{align*}

\begin{align*}
	\frac{\partial S}{\partial \beta}&=\sum_{t=k+1}^n 2\left(X_t-\mu-\alpha t-\beta(t-k)\right)(-(t-k)) \stackrel{\text { set }}{=} 0 \\
	\Rightarrow  \quad & \sum_{t=k+1}^n\left(X_t-\mu-\alpha t-\beta(t-k)\right)(t-k)=0\\
	& \sum_{t=1}^n t X_t=\mu \sum_{t=1}^n t+\alpha \sum_{t=1}^n t^2+\beta \sum_{t=k+1}^n(t-k) t \\
	\Rightarrow \quad & \sum_{t=k+1}^n(t-k) X_t=\mu \sum_{t=k+1}^n(t-k)+\alpha \sum_{t=k+1}^n t(t-k)+\beta \sum_{t=k+1}^n(t-k)^2.
\end{align*}

Therefore, we establish the normal equation set
\begin{align*}
	(A) \qquad &\sum_{t=1}^{n} X_{t}=n \hat{\mu}+\hat{\alpha} \sum_{t=1}^{n} t+\hat{\beta} \sum_{t=k+1}^{n}(t-k)\\  
	(B) \qquad &\sum_{t=1}^{n} t X_{t}=\hat{\mu} \sum_{t=1}^{n} t+\hat{\alpha} \sum_{t=1}^{n} t^{2}+\hat{\beta} \sum_{t=k+1}^{n}(t-k) t\\
	(C)  \qquad &\sum_{t=k+1}^{n}(t-k) X_{t}=\hat{\mu} \sum_{t=k+1}^{n}(t-k)+\hat{\alpha} \sum_{t=k+1}^{n} t(t-k)+\hat{\beta} \sum_{t=k+1}^{n}(t-k)^{2}
\end{align*}

For computational convenience, we define the follow notations
\begin{align*}
	a&=\sum_{t=1}^n t= \frac{n(n+1)}{2} \\
	b&=\sum_{t=k+1}^n(t-k)=\frac{(n-k)(n-k+1)}{2}\\
	c&=\sum_{t=1}^n t^2=\frac{n(n+1)(2 n+1)}{6} \\
	d&=\sum_{t=k+1}^n(t-k) t=\sum_{t-k+1}^n(t-k)^2+\sum_{t=k+1}^n k(t-k)\\
	&=(n-k)\frac{(n-k+1)[2(n-k)+1]}{6} +k \frac{(n-k)(n-k+1)}{2}\\
	&=\frac{(n-k)(n-k+1)(2n+k+1)}{6}\\
	e&=\sum_{t=k+1}^n(t-k)^2=\frac{(n-k)(n-k+1)(2(n-k)+1)}{6}\\
	V&_{1}=\sum_{t=1}^{n} X_{t}, \qquad 
	V_{2}=\sum_{t=1}^{n} t X_{t}, \qquad 
	V_{3}=\sum_{t=k+1}^{n}(t-k) X_{t}    
\end{align*}

The normal equation set becomes
\begin{align*}
	& V_{1}=n \hat{\mu}+a \hat{\alpha}+b \hat{\beta}  \tag{1}\\
	& V_{2}=a \hat{\mu}+c \hat{\alpha}+d \hat{\beta}  \tag{2}\\
	& V_{3}=b \hat{\mu}+d \hat{\alpha}+e \hat{\beta} \tag{3}
\end{align*}

From equation (1) we have $\hat{\mu}=\frac{V_{1}-a \hat{\alpha}-b \hat{\beta}}{n}$, and plug it into equations (2) and (3) to get

\begin{align*}
	& V_{2}-\frac{a V_{1}}{n}=\hat{\alpha}\left(c-\frac{a^{2}}{n}\right)+\hat{\beta}\left(d-\frac{a b}{n}\right) \qquad (2^{\prime}) \\
	& V_{3}-\frac{b V_{1}}{n}=\hat{\alpha}\left(d-\frac{a b}{n}\right)+\hat{\beta}\left(e-\frac{b^{2}}{n}\right) \qquad (3^{\prime})
\end{align*}

With $\left(2^{\prime}\right)$, we denote $\hat{\alpha}$ as:
\begin{align*}
	& \frac{V_{2}-\frac{a V_{1}}{n}-\hat{\beta}\left(d-\frac{a b}{n}\right)}{c-\frac{a^{2}}{n}}=\hat{\alpha} \\
	\Rightarrow \qquad & \frac{n V_{2}-a V_{1}-\hat{\beta}(n d-a b)}{n c-a^{2}}=\hat{\alpha}
\end{align*}

Plug $\hat{\alpha}$ into ($3^{\prime }$) to get
\begin{align*}
	&n V_{3}-b V_{1}=\hat{\alpha}(nd-a b)+\hat{\beta}\left(n e-b^{2}\right) \\  &n V_{3}-b V_{1}=  \left[ \frac{(n V_{2}-a V_{1})-\hat{\beta}(nd-ab)}{n c-a^{2}}\right](n d-a b)+\hat{\beta}(n e-b^{2})\\
	&(n V_{3}-b V_{1})(n c-a^{2})=\left[n V_{2}-a V_{1}-\hat{\beta}(n d-a b)\right](n d-a b)+\hat{\beta}(n e-b^{2})(n c-a^{2})\\
	&(n V_{3}-b V_{1})(n c-a^{2})-(n V_{2}-a V_{1} )(n d-a b)=
	\hat{\beta}\left[(n e-b^{2})(n c-a^{2})-(n d-a b)^{2}\right] \\
	& \Rightarrow \qquad \boxed{\hat{\beta}=\frac{(n V_{3}-b V_{1})(n c-a^{2})-(n V_{2}-a V_{1})(n d-a b)}{(n e-b^{2})(n c-a^{2})-(n d-a b)^{2}} }.
\end{align*}

Next, we compute the variances and covariances of $V_1,V_2$ and $V_3$: 
\begin{align*}
	\text{Var}\left(V_{1}\right)= & \text{Var}\left(\sum_{t=1}^{n} X_{t}\right)=\text{Var}\left(\sum_{t=1}^{n} \varepsilon_{t}\right)=n \sigma^{2} \\
	\text{Var}\left(V_{2}\right)= & \text{Var}\left(\sum_{t=1}^{n} t X_{t}\right)=\text{Var}\left(\sum_{t=1}^{n} t \varepsilon_{t}\right)=\sum_{t=1}^{n} \sigma^{2} t^{2} =  c\sigma^{2} \\
	\text{Var}\left(V_{3}\right)= & \text{Var}\left(\sum_{t=k+1}^{n}(t-k) X_{t}\right)=\sum_{t=k+1}^{n}(t-k)^{2} \sigma^{2} =e\sigma^{2}\\
	& =\sigma^{2}(n-k)(n-k+1)(2(n-k)+1)     
\end{align*}

\begin{align*}
	\text{Cov}\left(V_{1}, V_{2}\right)= & \text{Cov}\left(\sum_{t=1}^{n} X_{t}, \sum_{s=1}^{n} s X_{s}\right) =  \sum_{t=1}^{n} \sum_{s=1}^{n} s \text{Cov}\left(\varepsilon_{t}, \varepsilon_{s}\right) \\
	= & \sum_{t=1}^{n} t \sigma^{2}=\sigma^{2} \frac{n(n+1)}{2} = \boxed{a \sigma^{2}}\\
	\text{Cov}\left(V_{1}, V_{3}\right)= & \text{Cov}\left(\sum_{t=1}^{n} X_{t}, \sum_{s=k+1}^{n}(s-k) X_{s}\right) =  \sum_{t=k+1}^{n} \sum_{s=k+1}^{n} \text{Cov}\left(\varepsilon_{t},(s-k) \varepsilon_{s}\right) \\
	= & \sum_{t=k+1}^{n} \sum_{s=k+1}^{n}(s-k) \sigma^{2} \mathbbm{1}[s=t] \\
	= & \sigma^2 \sum_{t=k+1}^{n}(t-k)=\sigma^{2} \sum_{e=1}^{n-k} l=b \sigma^2    \\
	\text{Cov}\left(V_{2}, V_{3}\right) & =\text{Cov}\left(\sum_{t=1}^{n} t X_{t}, \sum_{s=k+1}^{n}(s-k) X_{s}\right) \\
	& =\sum_{t=k+1}^{n} t \sum_{s=k+1}^{n}(s-k) \text{Cov}\left(X_{t}, X_{s}\right) \\
	& =\sum_{t=k+1}^{n} t(t-k) \sigma^{2} =\sigma^{2}\left[(n-k)(n-k+1)\frac{2 n+k+1}{6}\right]=d\sigma^2
\end{align*}

$$
\hat{\beta}_k=\frac{\left(a V_{1}-n V_{2}\right)(a b-d n)-\left(b V_{1}-n V_{3}\right)\left(a^{2}-c n\right)}{(a b-d n)^{2}-\left(b^{2}-n e\right)\left(a^{2}-n c\right)}
$$

The numerator of $\beta$ is denoted as
\begin{align*}
	\text { Num } & =a(a b-d n) V_{1}-n(a b-d n) V_{2}-b\left(a^{2}-c n\right) V_{1}+n\left(a^{2}-c n\right) V_{3} \\
	& =\left(a^{2} b-a d n-b a^{2}+b c n\right) V_{1}+n(d n-a b) V_{2}+n\left(a^{2}-c n\right) V_{3} \\
	& =(b c-a d) n V_{1}+(d n-a b) n V_{2}+\left(a^{2}-c n\right) n V_{3},
\end{align*}

the denominator of $\beta$ is denoted as
\begin{align*}
	\text{Dem}=(a b-d n)^{2}-\left(b^{2}-n e\right)\left(a^{2}-n c\right),    
\end{align*}

and
\begin{align*}
	\hat{\beta}_k & =\frac{(b c-a d) n}{\text{Dem}} V_{1}+\frac{(d n-a b) n}{\text{Dem}} V_{2}+\frac{\left(a^{2}-c n\right) n}{\text{Dem}} V_{3} \\
	& =p_{1} V_{1}+p_{2} V_{2}+p_{3} V_{3}.
\end{align*}

\begin{align*}
	\text{Var}(\hat{\beta})&=\text{Var}\left(p_{1} V_{1}+p_{2} V_{2}+p_{3} V_{3}\right)\\
	& =p_{1}^{2} \text{Var}\left(\sum_{t=1}^{n} X_{t}\right)+p_{2}^{2} \text{Var}\left(\sum_{t=1}^{n} t X_{t}\right)+p_{3}^{2} \text{Var}\left(\sum_{t=k+1}^{n}(t-k) X_{t}\right) \\
	& +2 p_{1} p_{2} \text{Cov}\left(\sum_{t=1}^{n} X_{t}, \sum_{t=1}^{n} t X_{t}\right)+2 p_{1} p_{3} \text{Cov}\left(\sum_{t=1}^{n} X_{t}, \sum_{t=k+1}^{n}(t-k) X_{t}\right) \\
	& +2 p_{2} p_{3} \text{Cov}\left(\sum_{t=1}^{n} t X_{t}, \sum_{t=k+1}^{n}(t-k) X_{t}\right)\\
	= & p_{1}^{2} \cdot n \sigma^2+p_{2}^{2} \cdot c \sigma^2+p_{3}^{2} \cdot e \sigma^2 +2 p_{1} p_{2} \cdot a \sigma^2+2 p_{1} p_{3} \cdot b \sigma^2+2 p_{2} p_{3} \cdot d \sigma^2 \\
	= & \sigma^2\left[n p_{1}^{2}+c p_{2}^{2}+e p_{3}^{2}+2 a p_{1} p_{2}+2 b p_{1} p_{3}+2 d p_{2} p_{3}\right] \\
	=&\frac{\sigma^{2}}{\text{Dem}^{2}}  {\left[n^{3}(b c-a d)^{2}+c(d n-a b)^{2} n^{2}+e\left(a^{2}-c n\right)^{2} n^{2}\right.} \\
	&\qquad +2 a n^{2}(b c-a d)(d n-a b)+2 b n^{2}(b c-a d)\left(a^{2}-c n\right) \\
	&\qquad \left.+2 d n^{2}(d n-a b)\left(a^{2}-c n\right)\right]
\end{align*}

With \texttt{Mathematica}, we find
\begin{align*}
	\text{Var}(\hat{\beta}_k)&=\sigma^2
	\frac{V_n}{V_d}
\end{align*}
where
\begin{align*}
	V_n &=\frac{1}{864}\left(11 k^{2}-32 k^{3}+33 k^{4}-14 k^{5}+2 k^{6}\right) n^{5}\\
	&+\frac{1}{864}\left(-23 k+84 k^{2}-98 k^{3}+43 k^{4}-6 k^{5}\right) n^{6}\\
	&+\frac{1}{864}\left(12-72 k+91 k^{2}-12 k^{3}-27 k^{4}+14 k^{5}-2 k^{6}\right) n^{7}\\
	&+\frac{1}{864}\left(20-18 k-73 k^{2}+96 k^{3}-43 k^{4}+6 k^{5}\right) n^{8}\\
	&+\frac{1}{864}\left(-8+80 k-102 k^{2}+44 k^{3}-6 k^{4}\right) n^{9}\\
	&+\frac{1}{864}\left(-24+41 k-11 k^{2}+2 k^{3}\right) n^{10}\\
	&+\frac{1}{864}(-4-8 k) n^{11}\\
	&+\frac{1}{216}  n^{12},  
\end{align*}
and
\begin{align*}
	V_d& =\frac{\left(121 k^{4}-704 k^{5}+1750 k^{6}-2420 k^{7}+2029 k^{8}-1052 k^{9}+328 k^{10}-56 k^{11}+4 k^{12}\right) n^{4}}{5184}\\
	&+\frac{\left(-506 k^{3}+3320 k^{4}-9050 k^{5}+13406 k^{6}-11796 k^{7}+6302 k^{8}-1992 k^{9}+340 k^{10}-24 k^{11}\right) n^{5}}{5184}\\
	&+\frac{\left(793 k^{2}-6216 k^{3}+19208 k^{4}-31026 k^{5}+28848 k^{6}-15868 k^{7}+5061 k^{8}-860 k^{9}+60 k^{10}\right) n^{6}}{5184}\\
	&+\frac{\left(-552 k+5768 k^{2}-21322 k^{3}+38490 k^{4}-38010 k^{5}+21350 k^{6}-6788 k^{7}+1144 k^{8}-80 k^{9}\right) n^{7}}{5184}\\
	&+\frac{\left(144-2648 k+12966 k^{2}-27138 k^{3}+28332 k^{4}-15682 k^{5}+4782 k^{6}-800 k^{7}+60 k^{8}\right) n^{8}}{5184}\\
	&+\frac{\left(480-4048 k+10464 k^{2}-10940 k^{3}+4892 k^{4}-1116 k^{5}+196 k^{6}-24 k^{7}\right) n^{9}}{5184}\\
	&+\frac{\left(496-1840 k+1113 k^{2}+1082 k^{3}-715 k^{4}+100 k^{5}+4 k^{6}\right) n^{10}}{5184}\\
	&+\frac{\left(64+568 k-1384 k^{2}+512 k^{3}-80 k^{4}\right) n^{11}}{5184}\\
	&+\frac{\left(-144+392 k-24 k^{2}+16 k^{3}\right) n^{12}}{5184}\\
	&+\frac{(-32-64 k) n^{13}}{5184}\\
	&+\frac{n^{14}}{324}    
\end{align*}
Suppose $n\rightarrow \infty$, $k\rightarrow \infty$, and $\frac{k}{n}=t$. Dividing both the numerator and the denominator of $\text{Var}(\hat{\beta}_k)$ by $n^{16}$, we obtain
\begin{align*}
	\mbox{Var}(\hat{\beta}_k)=\frac{3}{n^3}\frac{\left(\frac{k}{n}\right)^3 \left(1-\frac{k}{n}\right)^3+o(n^{-1})}{\left(\frac{k}{n}\right)^6 \left(1-\frac{k}{n}\right)^6+o(n^{-1})}.    
\end{align*}

\newpage

To compute the covariance of  of $\hat{\beta}_{k}, \hat{\beta}_{l}$, $k<l$ for the jointpoint model, i.e., $\text{Cov}\left(\hat{\beta}_{k}, \hat{\beta}_{l}\right), k<l$, we first represent them as

\begin{align*}
	& \hat{\beta}_{k}=p_{1} V_{1}+p_{2} V_{2}+p_{3} V_{3} \\
	\text{where } & V_{1}=\sum_{t=1}^{n} X_{t}, \quad V_{2}=\sum_{t=1}^{n} t X_{t}, \quad V_{3}=\sum_{t=k+1}^{n}(t-k) X_{t} \\
	& \hat{\beta}_{l}=q_{1} W_{1}+q_{2} W_{2}+q_{3} W_{3} \\
	\text{where } & W_{1}=\sum_{t=1}^{n} X_{t}, \quad W_{2}=\sum_{t=1}^{n} t X_{t}, \quad W_{3}=\sum_{t=l+1}^{n}(t-l) X_{t} 
\end{align*}

Note $V_{1}=W_{1}, \quad V_{2}=W_{2}$. The coefficients $p_{1}, p_{2}, p_{3}, q_{1}, q_{2}, q_{3}$ depend on $k$, and $l$. Here,
\begin{align*}
	\text{Cov}\left(J_{k}, J_{l}\right)=\frac{\text{Cov}\left(\hat{\beta}_{k}, \hat{\beta}_{l}\right)}{\sqrt{\text{Var}\left(\hat{\beta}_{k}\right) \text{Var}\left(\hat{\beta}_{l}\right)} },  
\end{align*}

$\text{Var}\left(\hat{\beta}_{k}\right)$ and $\text{Var}\left(\hat{\beta}_{l}\right)$ were previously computed.

\begin{align*}
	\text{Cov}\left(\hat{\beta}_{k},\; \hat{\beta}_{l}\right) & =\text{Cov}\left(p_{1} V_{1}+p_{2} V_{2}+p_{3} V_{3}, q_{1} W_{1}+q_{2} W_{2}+q_{3} W_{3}\right) \\
	& =p_{1} q_{1} \underbrace{\text{Cov}\left(V_{1}, W_{1}\right)}_{(1)}+p_{1} q_{2} \underbrace{\text{Cov}\left(V_{1}, W_{2}\right)}_{(2)}+p_{1} q_{3} \underbrace{\text{Cov}\left(V_{1}, W_{3}\right)}_{(3)} \\
	& +p_{2} q_{1} \underbrace{\text{Cov}\left(V_{2}, W_{1}\right)}_{(4)}+p_{2} q_{2} \underbrace{\text{Cov}\left(V_{2}, W_{2}\right)}_{(5)}+p_{2} q_{3} \underbrace{\text{Cov}\left(V_{2}, W_{3}\right)}_{(6)} \\
	& +p_{3} q_{1} \underbrace{\text{Cov}\left(V_{3}, W_{1}\right)}_{(7)}+p_{3} q_{2} \underbrace{\text{Cov}\left(V_{3}, W_{2}\right)}_{(8)}+p_{3} q_{3} \underbrace{\text{Cov}\left(V_{3}, W_{3}\right)}_{(9)}.
\end{align*}

\begin{align*}
	(1)=\text{Cov}\left(V_{1}, W_{1}\right) & =\text{Cov}\left(V_{1}, W_{1}\right)=\text{Cov}\left(\sum_{t=1}^{n} X_{t}, \sum_{t=1}^{n} X_{t}\right) \\
	& =\text{Var}\left(\sum_{t=1}^{n} X_{t}\right)=n \sigma^{2}
\end{align*}

\begin{align*}
	(2)\text{Cov}\left(V_{1}, W_{2}\right) & =\text{Cov}\left(V_{1}, V_{2}\right)=\text{Cov}\left(\sum_{t=1}^{n} X_{t}, \sum_{t=1}^{n} t X_{t}\right) \\
	& =\sum_{t=1}^{n} t \sigma^{2}=\sigma^{2} \frac{n(n+1)}{2}=a \cdot \sigma^{2}
\end{align*}

\begin{align*}
	(3) &= \text{Cov}\left(V_{1}, W_{3}\right)=\text{Cov}\left(\sum_{t=1}^{n} X_{t}, \sum_{t=l+1}^{n}(t-l) X_{t}\right)\\
	& =\sigma^{2} \sum_{t=l+1}^{n}(t-l)=\sigma^{2} \frac{(n-l)(n-l+1)}{2}    
\end{align*}

\begin{align*}
	(4)&=\text{Cov}\left(V_{2}, W_{1}\right)=\text{Cov}\left(V_{2}, V_{1}\right)=\text{Cov}\left(V_{1}, V_{2}\right)=\text{Cov}\left(\sum_{t=1}^{n} X_{t}, \sum_{t=1}^{n} t X_{t}\right)\\
	&=\sigma^{2} \frac{n(n+1)}{2}=a \cdot \sigma^{2}    
\end{align*}

Note that (2) and (4) are the same.

\begin{align*}
	(5) &=\text{Cov}\left(V_{2}, W_{2}\right)=\text{Cov}\left(V_{2}, V_{2}\right)=\text{Var}\left(V_{2}\right)=\text{Var}\left(\sum_{t=1}^{n} t X_{t}\right)\\
	&=\sum_{t=1}^{n} t^{2} \sigma^{2}=\sigma^{2} \frac{n(n+1)(2 n+1)}{6}=c \cdot \sigma^{2}    
\end{align*}

\begin{align*}
	(6)&=\text{Cov}\left(V_{2}, W_{3}\right) =\text{Cov}\left(\sum_{t=1}^{n} t X_{t}, \sum_{t=l+1}^{n}(t-l) X_{t}\right) \\
	& =\sum_{t=l+1}^{n} t(t-l) \sigma^{2}=\sigma^{2} \frac{(n-l)(n-l+1)(2 n+l+1)}{6}
\end{align*}

\begin{align*}
	(7)&= \text{Cov}\left(V_{3}, W_{1}\right)=\text{Cov}\left(V_{3}, V_{1}\right)=\text{Cov}\left(V_{1}, V_{3}\right)=\text{Cov}\left(\sum_{t=1}^{n} X_{t}, \sum_{t=k+1}^{n}(t-k) X_{t}\right)\\
	&=\sigma^{2} \sum_{t=k+1}^{n}(t-k)=\sigma^{2} \frac{(n-k)(n-k+1)}{2}=b \cdot \sigma^{2}    
\end{align*}

\begin{align*}
	(8) & =\text{Cov}\left(V_{3}, W_{2}\right)=\text{Cov}\left(V_{3}, V_{2}\right)=\text{Cov}\left(V_{2}, V_{3}\right)=\text{Cov}\left(\sum_{t=1}^{n} t X_{t}, \sum_{t=k+1}^{n}(t-k) X_{t}\right) \\
	& =\sigma^{2} \sum_{t=k+1}^{n} t(t-k)=\sigma^{2} \frac{(n-k)(n-k+1)(2 n+k+1)}{6}=d \cdot \sigma^{2} 
\end{align*}

\begin{align*}
	(9) & =\text{Cov}\left(V_{3}, W_{3}\right)=\text{Cov}\left(\sum_{t=k+1}^{n}(t-k) X_{t}, \sum_{t=l+1}^{n}(t-l) X_{s}\right) \quad k<l \\
	& =\text{Cov}\left(\sum_{t=k+1}^{n}(t-k) X_{t}, \sum_{s=l+1}^{n}(s-l) X_{s}\right) \\
	& =\sum_{s=l+1}^{n}(s-k)(s-l) \text{Cov}\left(X_{s}, X_{s}\right)=\sigma^{2} \sum_{s=l+1}^{n}(s-k)(s-l) \\
	& =\sigma^{2} \sum_{s=l+1}^{n}(s-l+l-k)(s-l)- \\
	& =\sigma^{2} \sum_{s=l+1}^{n}\left[(s-l)^{2}+(l-k)(s-l)\right] \\
	& \left.=\sigma^{2}\left[\sum_{s=l+1}^{n}(s-l)^{2}+t l-k\right) \sum_{s=l+1}^{n}(s-l)\right] \\
	& =\sigma^{2}\left[\sum_{s=1}^{n-l} s^{2}+(l-k) \sum_{s=1}^{n-l} s\right] \\
	& =\sigma^{2}\left[\frac{(n-l)(n-l+1)(2 n-2 l+1)}{6}+(l-k) \frac{(n-l)(n-l+1)}{6}\right] \\
	& =\sigma^{2}[n-l)(n-l+1)\left[\frac{1}{6}(2 n-2 l+1)+\frac{1}{2}(l-k)\right]
\end{align*}

Here, we re-define the following constants (which were defined when computing the variance), but $b, d, e$ depend on the changepoint location $k$ and $l$

\begin{align*}
	& a=\sum_{t=1}^{n} t=\frac{1}{2} n(n+1) \\
	& b_{k}=\sum_{t=k+1}^{n}(t-k)=\frac{(n-k)(n-k+1)}{2}\\
	&b_{l}=\sum_{t=l+1}^{n}(t-l)=\frac{(n-l)(n-l+1)}{2} \\
	& c=\sum_{t=1}^{n} t^{2}=\frac{1}{6} n(n+1)(2 n+1) \\
	& d_k=\sum_{t=k+1}^{n}(t-k) t=\frac{1}{6}(n-k)(n-k+1)(2 n+k+1) \\
	& d_l=\sum_{t=l+1}^{n}(t-l) t=\frac{1}{6}(n-l)(n-l+1)(2 n+l+1) \\
	& e_{k}=\sum_{t=k+1}^{n}(t-k)^{2}=\frac{1}{6}(n-k)(n-k+1)(2 n-2 k+1) \\
	& e_{l}=\sum_{t=l+1}^{n}(t-l)^{2}=\frac{1}{6}(n-l)(n-l+1)(2 n-2 l+1)
\end{align*}

Combined with all these notations, we get
\begin{align*}
	\text{Cov}\left(\hat{\beta}_{k}, \hat{\beta}_{l}\right)& =p_{1} q_{1} \cdot(n 6)^{2}+p_{1} q_{2} \cdot a \sigma^{2}+p_{1} q_{3} \cdot \frac{1}{2}(n-l)(n-l+1) \sigma^{2} \\
	& +p_{2} q_{1} \cdot a \sigma^{2}+p_{2} q_{2} \cdot c \sigma^{2}+p_{2} q_{3} \cdot \frac{1}{6}(n-l)(n-l+1)(2 n+l+1) \sigma^{2} \\
	& +p_{3} q_{1} \cdot b \sigma^{2}+p_{3} q_{2} \cdot d \sigma^{2}+p_{3} q_{3} \cdot (n-l)(n-l+1)\left[\frac{1}{6}(2 n-2 l+1)+\frac{1}{2}(l-k)\right]\sigma^{2}
\end{align*}

Next, we need to deal $p_{i}^{\prime}$ s and $q_{i}$ 's. (We need to go back to the definition of $p_{i}^{\prime}$ s in the variance calculation)

Let $(*)$ denote the denominator for all $p_{i}$ 's at such that

\begin{align*}
	(*)=\left(a b_{k}-nd_{k}  \right)^{2}-\left(b_{k}^{2}-n e_{k}\right)\left(a^{2}-n c\right)    
\end{align*}


Let $(\Delta)$ denote the denominator for all $q_{i}$ 's at $l$.


\begin{align*}
	(\Delta)=(a b_l- nd_l )^{2}-\left(b l^{2}-n e_{l}\right)\left(a^{2}-n c\right) \quad .    
\end{align*}

so
\begin{align*}
	p_{1} q_{1}&= \frac{1}{(*)(\Delta)} \left(b_{k}  c-a d_{k}\right) n  \left(b_{l} c-a d_{l}\right) n \\   
	p_{1} q_{2}&=\frac{1}{(*)(\Delta)}\left(b_{k}  c-a d_{k}\right) n \left(d_{l}  n-a b l\right) n\\
	p_{1} q_{3}&=\frac{1}{(*)(\Delta)}\left(b_{k}  c-a d_{k}\right) n \left(a^{2}-c n\right) n\\
	p_{2} q_{1}&=\frac{1}{(*)(\Delta)}\left(d_{k} n-a b_{k}\right) n\left(b_{l} c-a d_{l}\right) n\\
	p_{2} q_{2}&=\frac{1}{(*)(\Delta)}\left(d_{k} n-a b_{k}\right) n\left(d_{l} n-ab_{l}\right) n\\
	p_{2} q_{3}&=\frac{1}{(*)(\Delta)}\left(d_{k} n-a b_{k}\right) n\left(a^{2}-c n\right) n\\
	p_{3} q_{1}&=\frac{1}{(*)(\Delta)}\left(a^{2}-c n\right) n(b_l c-a d_l) n\\
	p_{3} q_{2}&=\frac{1}{(*)(\Delta)}\left(a^{2}-c n\right) n(d_l n-a b_l) n\\
	p_{3} q_{3}&=\frac{1}{(*)(\Delta)}\left(a^{2}-c n\right) n\left(a^{2}-c n\right) n
\end{align*}

By substituting those notations into \texttt{Mathementica}, we find

\begin{align}
	\text{Cov}(\hat{\beta}_k, \hat{\beta}_l) = \frac{C_n}{C_d},    
\end{align}
where

\begingroup
\allowdisplaybreaks
\begin{align*}
	C_n&=\frac{1}{864}(11 k l-16 k^{2} l+5 k^{3} l-16 k l^{2}+23 k^{2} l^{2}-7 k^{3} l^{2}+5 k l^{3}-7 k^{2} l^{3}+2 k^{3} l^{3}) n^{5} \\
	& +\frac{1}{864}(-11 k+16 k^{2}-5 k^{3}-12 l+52 k l-43 k^{2} l+11 k^{3} l+16 l^{2}-46 k l^{2}+21 k^{2} l^{2}\\
	& \quad-3 k^{3} l^{2}-4 l^{3}+11 k l^{3}-3 k^{2} l^{3}) n^{6} \\
	& +\frac{1}{864}(12-36 k+20 k^{2}-4 k^{3}-36 l+51 k l-2 k^{2} l-5 k^{3} l+20 l^{2}-2 k l^{2}-17 k^{2} l^{2} \\
	& \quad+7 k^{3} l^{2}-4 l^{3}-5 k l^{3}+7 k^{2} l^{3}-2 k^{3} l^{3}) n^{7}\\
	&+\frac{1}{864}(20-10 k-12 k^{2}+6 k^{3}-8 l-49 k l+40 k^{2} l-11 k^{3} l-12 l^{2}+46 k l^{2}-21 k^{2} l^{2} \\
	&\quad+3 k^{3} l^{2}+4 l^{3}-11 k l^{3}+3 k^{2} l^{3}) n^{8} \\
	&+\frac{1}{864}(-8+40 k-20 k^{2}+4 k^{3}+40 l-62 k l+18 k^{2} l-20 l^{2}+18 k l^{2}-6 k^{2} l^{2}+4 l^{3}) n^{9} \\
	&+\frac{1}{864}(-24+21 k-4 k^{2}-k^{3}+20 l-3 k l+3 k^{2} l-4 l^{2}) n^{10} \\
	&+\frac{1}{864}(-4-4 k-4 l) n^{11} \\
	&+\frac{1}{216}n^{12}
\end{align*}
\endgroup

\begingroup
\allowdisplaybreaks
\begin{align*}
	C_d&=\frac{1}{5184}(121 k^{2} l^{2}-352 k^{3} l^{2}+363 k^{4} l^{2}-154 k^{5} l^{2}+22 k^{6} l^{2}-352 k^{2} l^{3}+1024 k^{3} l^{3} \\
	\quad&  -1056 k^{4} l^{3}+448 k^{5} l^{3}-64 k^{6} l^{3}+363 k^{2} l^{4}-1056 k^{3} l^{4}+1089 k^{4} l^{4} \\
	\quad&  -462 k^{5} l^{4}+66 k^{6} l^{4}-154 k^{2} l^{5}+448 k^{3} l^{5}-462 k^{4} l^{5}+196 k^{5} l^{5}-28 k^{6} l^{5} \\
	\quad &  +22 k^{2} l^{6}-64 k^{3} l^{6}+66 k^{4} l^{6}-28 k^{5} l^{6}+4 k^{6} l^{6}) n^{4} \\
	& +\frac{1}{5184}(-253 k^{2} l+736 k^{3} l-759 k^{4} l+322 k^{5} l-46 k^{6} l-253 k l^{2}+1848 k^{2} l^{2}-3766 k^{3} l^{2} \\
	\quad& +3245 k^{4} l^{2}-1242 k^{5} l^{2}+168 k^{6} l^{2}+736 k l^{3}-3766 k^{2} l^{3}+6272 k^{3} l^{3} \\
	\quad&  -4610 k^{4} l^{3}+1564 k^{5} l^{3}-196 k^{6} l^{3}-759 k l^{4}+3245 k^{2} l^{4}-4610 k^{3} l^{4} \\
	& \quad +2838 k^{4} l^{4}-800 k^{5} l^{4}+86 k^{6} l^{4}+322 k l^{5}-1242 k^{2} l^{5}+1564 k^{3} l^{5} \\
	\quad&  -800 k^{4} l^{5}+168 k^{5} l^{5}-12 k^{6} l^{5}-46 k l^{6}+168 k^{2} l^{6}-196 k^{3} l^{6}+86 k^{4} l^{6} \\
	\quad&  -12 k^{5} l^{6}) n^{5} \\
	& +\frac{1}{5184}(132 k^{2}-384 k^{3}+396 k^{4}-168 k^{5}+24 k^{6}+529 k l-2724 k^{2} l+4558 k^{3} l \\
	\quad&  -3365 k^{4} l+1146 k^{5} l-144 k^{6} l+132 l^{2}-2724 k l^{2}+9300 k^{2} l^{2} \\
	\quad& -11980 k^{3} l^{2}+7044 k^{4} l^{2}-1932 k^{5} l^{2}+204 k^{6} l^{2}-384 l^{3}+4558 k l^{3} \\
	\quad&  -11980 k^{2} l^{3}+12420 k^{3} l^{3}-5858 k^{4} l^{3}+1204 k^{5} l^{3}-88 k^{6} l^{3}+396 l^{4} \\
	\quad& -3365 k l^{4}+7044 k^{2} l^{4}-5858 k^{3} l^{4}+2245 k^{4} l^{4}-342 k^{5} l^{4}+12 k^{6} l^{4}-168 l^{5} \\
	\quad& +1146 k l^{5}-1932 k^{2} l^{5}+1204 k^{3} l^{5}-342 k^{4} l^{5}+36 k^{5} l^{5}+24 l^{6}-144 k l^{6} \\
	\quad& +204 k^{2} l^{6}-88 k^{3} l^{6}+12 k^{4} l^{6}) n^{6} \\
	& +\frac{1}{5184}(-276 k+1228 k^{2}-1816 k^{3}+1176 k^{4}-352 k^{5}+40 k^{6}-276 l+3312 k l \\
	\quad& -8845 k^{2} l+9380 k^{3} l-4587 k^{4} l+1006 k^{5} l-82 k^{6} l+1228 l^{2}-8845 k l^{2} \\
	\quad& +17378 k^{2} l^{2}-14066 k^{3} l^{2}+5253 k^{4} l^{2}-766 k^{5} l^{2}+22 k^{6} l^{2}-1816 l^{3} \\
	\quad& +9380 k l^{3}-14066 k^{2} l^{3}+8752 k^{3} l^{3}-2546 k^{4} l^{3}+292 k^{5} l^{3}-4 k^{6} l^{3} \\
	\quad& +1176 l^{4}-4587 k l^{4}+5253 k^{2} l^{4}-2546 k^{3} l^{4}+516 k^{4} l^{4}-36 k^{5} l^{4}-352 l^{5} \\
	\quad& +1006 k l^{5}-766 k^{2} l^{5}+292 k^{3} l^{5}-36 k^{4} l^{5}+40 l^{6}-82 k l^{6}+22 k^{2} l^{6} \\
	\quad& -4 k^{3} l^{6} ) n^{7} \\
	& +\frac{1}{5184} (144-1324 k+2948 k^2-2616 k^3+1064 k^4-176 k^5+8 k^6-1324 l+7070 k l \\
	\quad & -10953 k^{2} l+6976 k^{3} l-1931 k^{4} l+134 k^{5} l+16 k^{6} l+2948 l^{2}-10953 k l^{2} \\
	\quad& +12252 k^{2} l^{2}-5734 k^{3} l^{2}+1085 k^{4} l^{2}-66 k^{5} l^{2}-2616 l^{3}+6976 k l^{3} \\
	\quad& -5734 k^{2} l^{3}+2328 k^{3} l^{3}-350 k^{4} l^{3}+12 k^{5} l^{3}+1064 l^{4}-1931 k l^{4} \\
	\quad&  +1085 k^{2} l^{4}-350 k^{3} l^{4}+36 k^{4} l^{4}-176 l^{5}+134 k l^{5}-66 k^{2} l^{5}+12 k^{3} l^{5}+8 l^{6} \\
	\quad&  +16 k l^{6}) n^{8}\\
	& +\frac{1}{5184}(480-2024 k+2464 k^{2}-1168 k^{3}+160 k^{4}+32 k^{5}-8 k^{6}-2024 l+5536 k l \\
	\quad & -4302 k^{2} l+1164 k^{3} l+98 k^{4} l-48 k^{5} l+2464 l^{2}-4302 k l^{2}+2244 k^{2} l^{2} \\
	\quad & -688 k^{3} l^{2}+66 k^{4} l^{2}-1168 l^{3}+1164 k l^{3}-688 k^{2} l^{3}+176 k^{3} l^{3}-12 k^{4} l^{3} \\
	\quad & +160 l^{4}+98 k l^{4}+66 k^{2} l^{4}-12 k^{3} l^{4}+32 l^{5}-48 k l^{5}-8 l^{6}) n^{9} \\
	\quad & +\frac{1}{5184}(496-920 k+292 k^{2}+176 k^{3}-148 k^{4}+24 k^{5}-920 l+529 k l+365 k^{2} l  \\
	\quad & -270 k^{3} l+48 k^{4} l+292 l^{2}+365 k l^{2}+121 k^{2} l^{2}-22 k^{3} l^{2}+176 l^{3}-270 k l^{3} \\
	\quad & -22 k^{2} l^{3}+4 k^{3} l^{3}-148 l^{4}+48 k l^{4}+24 l^{5}) n^{10} \\
	& +\frac{1}{5184}(64+284 k-364 k^{2}+168 k^{3}-24 k^{4}+284 l-656 k l+88 k^{2} l-16 k^{3} l-364 l^{2}\\
	&\quad +88 k l^{2}+168 l^{3}-16 k l^{3}-24 l^{4}) n^{11} \\
	&+\frac{1}{5184}(-144+196 k-44 k^{2}+8 k^{3}+196 l+64 k l-44 l^{2}+8 l^{3}) n^{12} \\
	& +\frac{1}{5184}(-32-32 k-32 l) n^{13} \\
	&+\frac{1}{324}n^{14}
\end{align*}
\endgroup


\newpage

\bibliographystyle{plainnat}
\bibliography{arxiv.bib}

\end{document}